\documentclass[epj]{svjour}
\usepackage{t1enc}
\usepackage[utf8]{inputenc}

\usepackage{graphicx}   
\usepackage{epsfig}
\usepackage{epstopdf}

\begin{document}
\title{Percolation in the classical blockmodel}
\author{Maksymilian Bujok \and Piotr Fronczak \and Agata Fronczak }
\institute{Faculty of Physics, Warsaw University of Technology, Koszykowa 75, PL-00-662 Warsaw, Poland}
\date{Received: \today / Revised version: date}
%
\abstract{Classical blockmodel is known as the simplest among models of networks with community structure. The model can be also seen as an extremely simply example of interconnected networks. For this reason, it is surprising that the percolation transition in the classical blockmodel has not been examined so far, although the phenomenon has been studied in a variety of much more complicated models of interconnected and multiplex networks. In this paper we derive the self-consistent equation for the size the global percolation cluster in the classical blockmodel. We also find the condition for percolation threshold which characterizes the emergence of the giant component. We show that the discussed percolation phenomenon may cause unexpected problems in a simple optimization process of the multilevel network construction. Numerical simulations confirm the correctness of our theoretical derivations.
\PACS{
      {89.75.Fb}{Structures and organization in complex systems}   \and
      {64.60.aq}{Networks} \and 
      {64.60.ah}{Percolation}
     } 
} 
\maketitle
\section{Introduction}
\label{sec1}
For over a decade scientists of various disciplines have been showing an increasing interest in the field of complex networks \cite{bookDorogov,bookBarrat,bookNewman}. It was related to the rapid development of the Internet, which in turn has made available a huge amount of data on the structure and functioning of many real networks, such as social networks, biological networks (e.g. food chains) and data communication networks (e.g. Internet), among others.

Initially, in the awareness of researchers, complex networks functioned as isolated systems. The  structural and functional properties of individual networks were investigated. Only recently studying of multiplex network systems has begun in which individual networks may interact with each other \cite{2010Buldyrev,2012Gao}. An example of such interacting networks is a network that supplies energy to the computer network which in turn controls the energy distribution in the first network.

To understand the functioning of both types of networks (i.e. single networks and systems of interacting networks) the corresponding models are created and different dynamic processes are defined, such as the spread of epidemics and the opinion formation or diffusion processes. In many of these processes, the underlying phenomenon is percolation. For example, using percolation theory it has become possible to understand, why real networks are highly robust to random failures but fragile against attacks \cite{2000aCohen,2000bCohen}. Recently, percolation theory was also used in a discussion on structural properties of the important class of  multiplex network systems, see e.g. \cite{2010ParshaniPRL,2011GaoPRL,2012EPL,2014Bianconi}.

In this work we study the phenomenon of percolation in the so-called \emph{classical blockmodel} which has a long tradition of research in both social and computer sciences~\cite{1983Holland,1992Faust,1992Anderson,1997Snijders,2008Airoldi,2009Goldenberg} and recently also in the study of complex networks~\cite{2002GirvanPNAS,2004NewmanPRE,2008LanciPRE,2011KarrerPRE,2013FronczakPRE}. The classical blockmodel was introduced by Holland, Laskey and Leinhardt in 1983~\cite{1983Holland}. This model is interesting from the point of view of recent studies because it can be seen as a model of a single network with community structure~\cite{2010Fortunato,2012NewmanNature}, but also as a simple model of a two-level network which consists of smaller networks (i.e. network of networks)~\cite{2010Buldyrev,2012Gao}. In such a two-level network, nodes may be connected through local edges (at the first level) and global ones (at the second level). This model can also be regarded as a generalization of the classical random graph of Erd\"{o}s--R\'{e}nyi (ER) wherein each of the $N$ nodes is assigned to one of $K$ blocks (communities, local area networks) of the same size. The probability of the existence of an edge is different for the nodes belonging to the same block (local level) and different, when nodes belong to different blocks (global level).

Further in the paper we deal with the following issues: In section \ref{sec2} we recall a simple microscopic formalism described in our earlier work \cite{2005FronczakAIP}, which allowed us to calculate the percolation threshold and the size of the largest connected component in classical random graphs. This formalism, as one of many that have been used to describe the phenomenon of percolation in classical random graphs (see e.g.~\cite{1959Erdos,1998Molloy,2000Callaway}), is in our opinion the simplest one. In section \ref{sec3} we use this formalism to study percolation phase transition in the classical blockmodel. We determine the percolation threshold and the size of the largest connected component (i.e. percolation cluster). These results are compared with the results of numerical simulations. In section \ref{sec4}, we discuss a simple optimization process of the construction procedure of two-level  distribution networks. The procedure is based on the results obtained in section \ref{sec3}. Section \ref{sec5} is devoted to the summary of results.

\section{Percolation in classical random graphs}\label{sec2}

Classical random graphs were first discussed by Erd\"{o}s and R\'{e}nyi in the articles of the mid-twentieth century. In this paper, the term \emph{classical random graph} refers to a certain generalisation of the original ER model: Classical random graph means the graph of $N$ numbered vertices (nodes), where each pair of vertices is connected by an edge with probability $p$.

In this section, our goal is to remind one of the method \cite {2005FronczakAIP}, that allows to determine the size, $N_G$, of the largest cluster (i.e. the number of nodes belonging to the cluster) in classical random graphs. From previous works on this subject, we know that for $N\gg 1$, when the average node degree is less than one, $\langle k\rangle<1$, the relative size of the largest cluster, $S=N_G/N$, is equal to zero. Only for $\langle k\rangle>1$, the parameter $S$ becomes greater than zero and increases, reaching the value of $S=1$ when all nodes belong to the same cluster. The critical value of $\langle k \rangle = 1 $ is called the percolation threshold and the largest connected component for $\langle k \rangle> 1$ is called the percolation cluster. In percolation theory, the relative size of the percolation cluster, $S$, which expresses the probability that a randomly selected node of the graph belongs to that cluster, acts as an order parameter of percolation phase transition.

To determine the value of parameter  $S$, let us consider a randomly selected node $i$ belonging to the ER graph. From the design procedure it follows that  in the classical random graph all nodes and all edges are, from a statistical point of view, the same. This means that probability $S$, that the randomly chosen vertex $i$ belongs to the largest cluster is equal to probability that at least one of $N-1$ other vertices of the graph belongs to such a cluster and it is connected to the node $i$. Thus, if $A_{\left\{i,j\right\}}$ corresponds to the event that the edge $\left\{i,j\right\}$ (between the considered node $i$ and a node $j$) belongs to the largest cluster, then the probability that the node $i$ also belongs to this cluster becomes:
\begin{equation}
\label{eq:equation2}
S=P \left( \bigcup_{j=1}^{N-1} A_{\left\{ i,j \right\}} \right).
\end{equation}
Since the events $A_{\left\{i,j\right\}}$ are independent, then using the known theorem on the  sum of independent events~\cite{2005FronczakAIP,2004FronczakPRE}, i.e.
\begin{equation}\label{lemat}
P\left(\bigcup_{j=1}^{N-1} A_{\left\{ i,j \right\}}\right)=1-\exp \left[ -\sum_{j=1}^{N-1} P(A_{\left\{ i,j \right\}}) \right],
\end{equation}
Eq.~(\ref{eq:equation2}) can be written as:
\begin{equation}
\label{eq:equation3}
S=1-\exp\left[-\sum_{j=1}^{N-1} P(A_{\left\{i,j\right\}})\right].
\end{equation}

\begin{figure}
\resizebox{0.95\columnwidth}{!}{%
\includegraphics{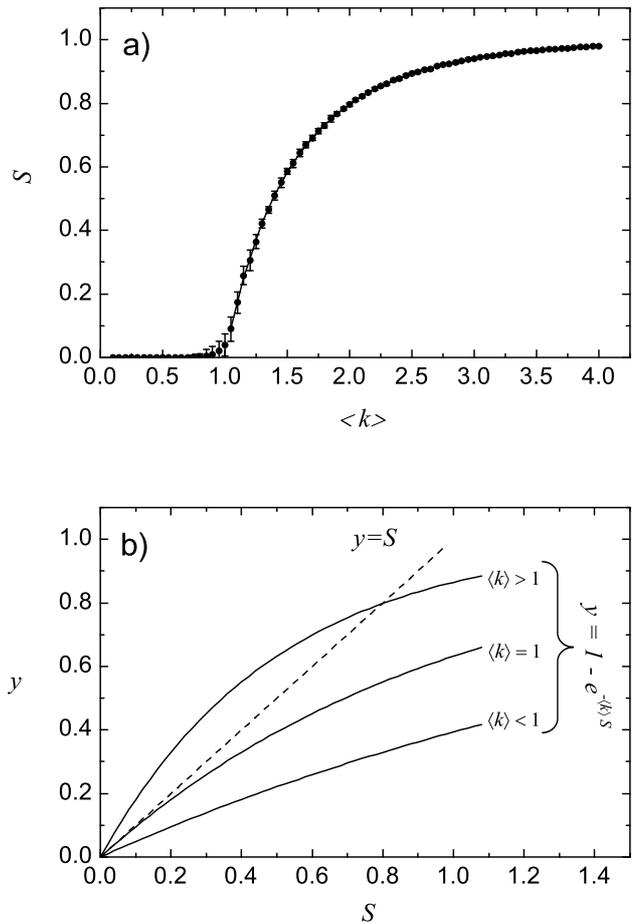}}
\caption{\textbf{Percolation in classical random graphs.} a) Size of the giant component $S$ versus the average degree $\langle k\rangle$ (the scattered points are results of numerical simulations and the solid line is theoretical solution obtained from Eq.~(\ref{eq6})). b) Graphical solution of Eq.~(\ref{eq6}) for the size of the giant component (detailed description is given in the text).} \label{fig1}
\end{figure}

Eq.~(\ref{eq:equation3}) can be further simplified, using the fact that $P(A_{\left\{i,j\right\}})$ is equal to the product of probability $p$ which means that the edge $\{i,j\}$ exists and probability $S$ that $j$ belongs to the largest connected component:
\begin{equation}
\label{eq:equation4}
P(A_{\left\{ i,j \right\}})=pS.
\end{equation}
Therefore, substituting the expression (\ref{eq:equation4}) to equation (\ref{eq:equation3}) we obtain the well-known self-consistency equation for the size of percolation cluster:
\begin{equation}
\label{eq6}
S=1-e^{-\langle k\rangle S},
\end{equation}
where  $\langle k\rangle=p(N-1)\simeq pN$ is the average degree of nodes in in the classical random graph. 

For the first time, the equation (\ref{eq6}) was given by Erd\"os and R\'enyi in 1959 \cite{1959Erdos}. It gives the relative size of the giant component for any given value of the mean degree $\langle k\rangle$. However, although this equation is very simple it does not have a simple solution for the size of the giant component as a function of $\langle k\rangle$ in closed form. The numerical solution of Eq.~(\ref{eq6}) for $S$ as compared with results of Monte Carlo simulations is shown in Fig.~\ref{fig1}a. Fortunately, graphical solution of this equation, which is given in Fig.~\ref{fig1}b, in a very suggestive way illustrates the percolation transition in classical random graphs (see also Fig.~12.1 in \cite{bookNewman} and Fig.~3 in \cite{2005FronczakAIP}). The three solid curves shown in Fig.~\ref{fig1}b represent the right hand side of Eq.~(\ref{eq6}), i.e. the function $y(S)=1-e^{-\langle k\rangle S}$, for different values of $\langle k\rangle$. The dashed line in the figure is the linear function $y(S)=S$. Where the line and the curve cross, the corresponding value of $S$ is a solution to Eq.~(\ref{eq6}).

As the figure shows, depending on the value of $\langle k\rangle$ there may be either one solution for $S$ or two. For small $\langle k\rangle$ (bottom curve in Fig.~\ref{fig1}b) the only solution is $S=0$, which means that the percolation cluster does not exists. For sufficiently large $\langle k\rangle$ (top curve), the equation (\ref{eq6}) has two solutions, $S=0$ and $S\neq 0$, the first of which is unstable. It means, that the percolation cluster of size $S\neq 0$ appears in a graph. The middle curve in the figure corresponds to the transition point between the two regimes. The characteristic point (the percolation threshold) is where the gradient of the curve $y(S)=1-e^{-\langle k\rangle S}$ and the slope of the dashed line $y(S)=S$ match at $S=0$. Therefore, the point can be determined from the first derivative of the right-hand side of Eq.~(\ref{eq6}):
\begin{equation}\label{eq6a}
\left.\frac{d(1-e^{-\langle k\rangle S})}{dS}\right|_{S=0}=1.
\end{equation}
From the above equation one immediately finds that the percolation threshold in classical random graphs is given by:
\begin{equation}\label{eq6b}
\langle k\rangle=1.
\end{equation}

\section{Percolation in the classical blockmodel}\label{sec3}

\begin{figure}
\centerline{\resizebox{0.9\columnwidth}{!}{%
  \includegraphics{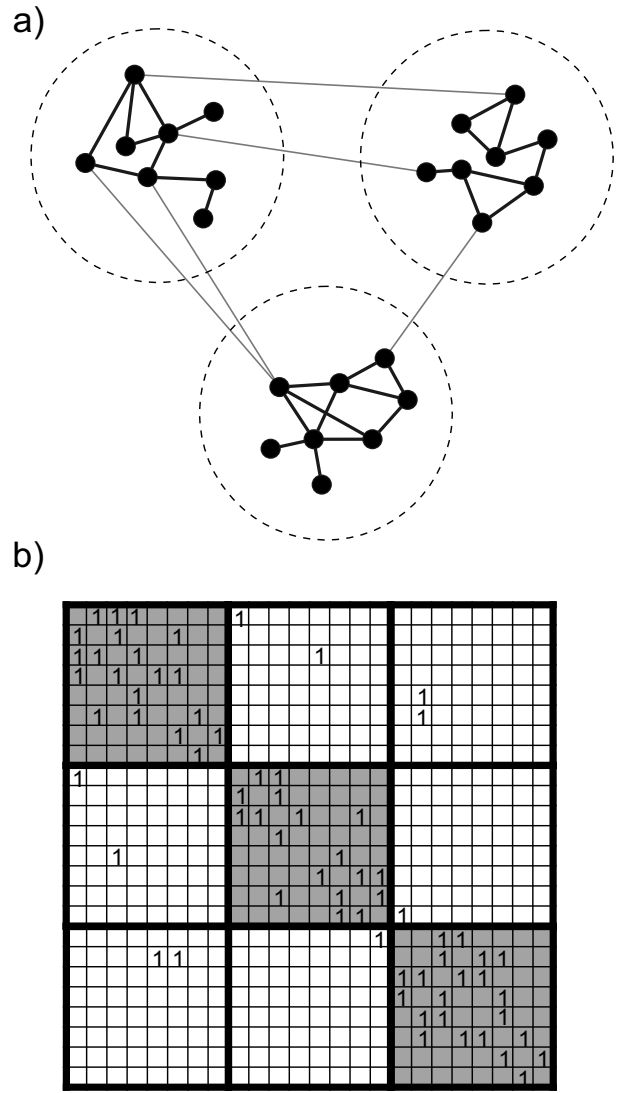}}}
\caption{\textbf{Classical blockmodel.} a) A small network with block structure of the type considered in this paper. In this case, there are three blocks $K=3$ of size $R=8$, denoted by the dashed circles, which have dense local connections. Global connections between the blocks are sparse. b) The adjacency matrix of the graph shown in Fig.~\ref{fig1}a. Gray areas along the diagonal represent adjacency matrices for the nodes belonging to the same blocks. Non zero matrix elements occurring outside the gray areas represent global connections.} \label{fig2}
\end{figure}

In the classical blockmodel (see Fig.~\ref{fig2}), each of $N$ nodes is assigned to one of $K$ blocks of the same size $R$, i.e. $N=KR$. In the model, blocks (modules) are ER graphs. The probability $p$ that there is an edge between the nodes belonging to the same block is usually different from the probability $q$ that there is an edge between the nodes belonging to different blocks. Further in this paper, the edges connecting the nodes that belong to the same block will be called \emph{local} connections while the edges connecting nodes from different blocks will be referred to as \emph{global} ones. Similarly, when we talk about percolation inside the blocks and at the level of local connections we will use the notion of the \emph{local percolation cluster}. The \emph{global percolation cluster} will be called the largest connected component of the whole graph. Such a cluster will be built with the nodes belonging to different blocks, which are connected through both local or global edges.

In this section our aim is to find the expression for percolation threshold and to calculate the relative size of the global percolation cluster in the classical blockmodel. In order to do it we will use a method similar to that of the previous section which allowed us to describe the percolation transition in classical random graphs.

Thus, let $G$ be the probability that a randomly chosen node $i$ in the classical blockmodel belongs to the global percolation cluster. When we take into account only local connections of this node, there are only two possible cases: a) the node $i$ belongs to the local percolation cluster of size $S$,  and b) the node belongs to one of the smaller local clusters of size $s$, wherein the probability that a randomly chosen vertex belongs to a component of size $s$ is given by (see Eq.~(12.50) in~\cite{bookNewman}):
\begin{equation}\label{pis}
\pi_s=\frac{e^{-\langle k\rangle s}(\langle k\rangle s)^{s-1}}{s!},
\end{equation}
where
\begin{equation}
\sum_{s}\pi_s=1-S,
\end{equation}
and
\begin{equation}\label{sk1}
\langle k\rangle=pR
\end{equation}
is the mean \emph{local degree} of nodes in the classical blockmodel \footnote{cf. the so-called \emph{internal} and \emph{external} node degrees in blockmodels, as they were defined in \cite{2013FronczakPRE};}. In both the cases, the probability that the node $i$ belongs to the global percolation cluster is equal to the probability that node $i$ itself or at least one of the nodes (e.g., node $j$) which belong to the same local cluster (percolation or not) are connected by a global edge to a node belonging to the global percolation cluster.

Therefore, let $B_{jl}$ be the event that the global edge $\{j,l\}$ between the nodes $j$ and $l$ belongs to the global percolation cluster (the node $j$ should be thought of as a node belonging to the same block as well as to the same local cluster as the node $i$). It follows that the parameter $G$, which is equal to the probability that a randomly selected node $ $ is a global percolation cluster is equal to:
\begin{equation}\label{eq7}
G=SP\left( \bigcup_{j=1}^{SR}\bigcup_{l=1}^{N-R} B_{\left\{j,l\right\}}\right)+\sum_{s}\pi_sP\left(\bigcup_{j=1}^{s} \bigcup_{l=1}^{N-R}B_{\left\{j,l\right\}}\right),
\end{equation}
where the summation over $j$ is carried out over all the nodes belonging to the same local cluster to which the node $i$ belongs while the summation over $l$ is carried out over the nodes belonging to other modules.

Then, using the theorem on the sum of independent events, cf. Eq.~(\ref{lemat}), and proceeding similarly as in Sec.~\ref{sec2}, the expression (\ref{eq7}) can be simplified to:
\begin{eqnarray}\nonumber
G&=&S\left(1-\exp \left[ -\sum_{j=1}^{SR}\sum_{l=1}^{N-R}P(B_{\{j,l\}}) \right]\right)+\\\label{eq8}&&\sum_s\pi_i\left(1-\exp \left[ -\sum_{j=1}^{s}\sum_{l=1}^{N-R}P(B_{\{j,l\}})\right]\right),
\end{eqnarray}
where
\begin{equation}\label{eq9}
P(B_{\{j,l\}})=qG,
\end{equation}
is the probability that a node $l$ at the end of the global edge $\{j,l\}$  belongs to the global percolation cluster. Finally, substituting Eq.~(\ref{eq9}) to (\ref{eq8}), we get the self-consistent equation for $G$:
\begin{eqnarray}\nonumber
G&=&S\left(1-e^{-qSR(N-R)G}\right)+\\&&\sum_s\pi_s\left(1-e^{-qs(N-R)G}\right)=f(G)
\label{fodG}.
\end{eqnarray}

\begin{figure}
\centerline{\resizebox{\columnwidth}{!}{%
\includegraphics{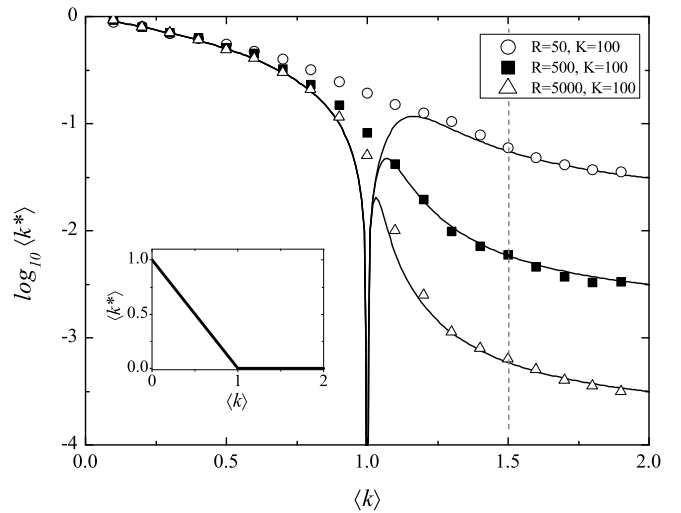}}}
\caption{\textbf{Phase diagram for percolation transition in the classical blockmodel.} Points placed in the figure represent results of numerical simulations for three different network sizes and correspond to values of the parameters $\langle k\rangle$ and $\langle k^*\rangle$ for which the largest cluster (averaged over 10 network realizations) starts to form. Solid lines correspond to Eq.~(\ref{eq12}).} \label{fig3}
\end{figure}

The right hand side of Eq.~(\ref{fodG}), i.e. the function $f(G)$, is continuous and monotonically increasing for $G$ in the range from $0$ to $1$. In fact, the behavior of this function is very similar to the behavior of the right hand side of Eq.~(\ref{eq6}) which was discussed in Sec.~\ref{sec2}. In particular, it is easy to see that the value of $G=0$ (which means lack of the global percolation cluster) is always the solution of Eq.~(\ref{fodG}). Furthermore, for certain values of  $p,q,R,K$ parameters, a non-zero solution, $G\neq 0$, of this equation appears which characterizes the emergence of the global percolation cluster.

Thus, similarly as in classical random graphs, the condition for the emergence of the global percolation cluster in the classical blockmodel may be obtained by comparing the first derivatives on both sides of Eq.~(\ref{fodG}) at $G=0$. After simple algebra the condition for the global percolation transition, i.e. the point at which a sample-spanning global cluster first appears, takes the following form:
\begin{equation}\label{eq12}
\langle k^*\rangle=\frac{1}{S^2R+(1-S)\langle s\rangle},
\end{equation}
where
\begin{equation}\label{kq}
\langle k^*\rangle=q(N-R),
\end{equation}
is the mean number of global connections attached to a node, i.e. the average \emph{global degree} of nodes in the classical blockmodel, whereas $\langle s\rangle$ is the mean size of the local component to which a randomly chosen node belongs (see Eq.~(12.34) in \cite{bookNewman}),
\begin{equation}\label{sk}
\langle s\rangle=\frac{\sum_ss\pi_s}{\sum_s\pi_s}= \frac{1}{(1-S)(1-\langle k\rangle+\langle k\rangle S)}.
\end{equation}

\begin{figure}
\centerline{\resizebox{\columnwidth}{!}{%
\includegraphics{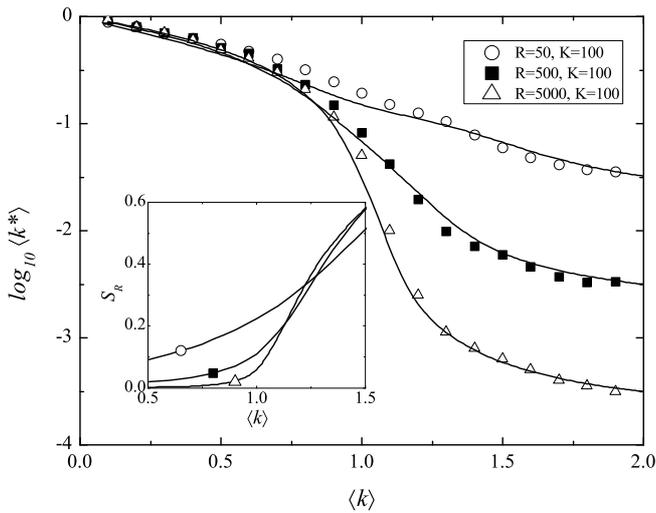}}}
\caption{\textbf{Percolation threshold in the finite-size classical blockmodel.} Symbols correspond to the same numerical results as shown in Fig.~\ref{fig3}. Solid lines represent Eq.~(\ref{eq12}), in which the parameter $S$ characterizing the size of the giant component in infinite networks was replaced by the parameter $S_R$, which applies to finite networks. In the inset, the numerical vaules of $S_R$ for different network sizes (as indicated by the respective symbols) are shown.} \label{fig5}
\end{figure}

Figure \ref{fig3} shows the results of numerical simulations for percolation threshold in the classical blockmodel against the theoretical curves predicted by Eq.~(\ref{eq12}). Notable is that while the theoretical and numerical results agree for $\langle k\rangle<1$ and $\langle k\rangle>1$, for $\langle k\rangle$ close to $1$ the theoretical curves diverge, which is not in agreement with numerical data. This discrepancy is due to the fact that Eq.~(\ref{sk}) for $\langle s\rangle$ diverges at $\langle k\rangle=1$. At this point, $S=0$ and the denominator of Eq.~(\ref{sk}) vanishes. However, this is only true for infinite graphs. In finite ER graphs, the size of the largest cluster $S_R>0$ for $\langle k\rangle=1$ (see inset in Fig.~\ref{fig5}). If we replace $S$ in Eq.~(\ref{eq12}) with values of $S_R$ obtained from numerical simulations we get rid of the divergence problem (see Fig.~\ref{fig5}).

\begin{figure}
\centerline{\includegraphics[width=\columnwidth]{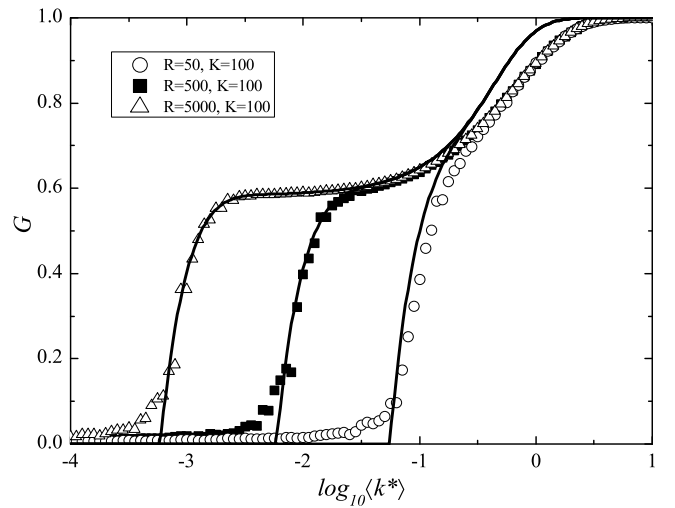}}
\caption{\textbf{Percolation in the classical blockmodel.} Size of the global percolation cluster $G$ versus the average global degree $\langle k^*\rangle$ for the fixed value of $\langle k\rangle=1.5$. The solid lines represent theoretical solution of Eq.~(\ref{MFfodG}) for $G$.}\label{fig4}
\end{figure}

In Fig.~\ref{fig3}, the dashed line indicates the fixed value of the parameter $\langle k\rangle=1.5$ for which the relative size of the global percolation cluster, $G$, is examined against the theoretical prediction of the self-consistent mean-field version of Eq.~(\ref{fodG}):
\begin{equation}\label{MFfodG}
G\simeq S\left(1-e^{-SR\langle k^*\rangle G}\right)+(1-S)\left(1-e^{-\langle s\rangle\langle k^*\rangle G}\right).
\end{equation}
Figure~\ref{fig4} shows comparison between the theoretical curves given by Eq.~(\ref{MFfodG}) and the numerically obtained values of $G$ as a function of $\langle k^*\rangle$ for three different network sizes. The numerical simulations provide an independent test for correctness of our theoretical calculations.

\section{Network optimization procedure}\label{sec4}

In this section we discuss a possible application of the discussed percolation phenomenon, namely a network optimization process which has been widely studied in recent years \cite{epjb_gastner,book_sole,book_schweitzer}. Such an optimization is of common interest in many different areas, among them electrical engineering, telecommunication, road construction and trade logistics. In the rest of this section we will concentrate on the economic optimization process which typically aims to reduce costs while increasing revenues.

Thus, let us consider a business entity such as a
network operator or a network manager who earns profit from connecting local or regional networks. Its revenue is related to the number in interconnected customers, i.e. it is proportional to the size of the giant connected component $G$, while the costs arise from  constructing and maintaining links between regional networks, i.e. they depend on the number of links $L=\langle k^*\rangle N$ and this dependence is, according to the so-called economies-of-scale effects, sublinear. Such a sublinear behavior arises from the fact that the decrease in unit cost of a product or service results from large-scale operations and the fixed costs are spread out over more
units of output \cite{economyofscale}. In the following, for the purpose of better demonstration, we assume that the cost scales logarithmically with $L$. Taking above into consideration, one has to optimize the following equation:
\begin{equation}\label{eq:cost}
C=\left( 1 - \lambda \right) \log(L) - \lambda G,
\end{equation}
where $\lambda$ is a parameter controlling a ratio between expenses and income parts. The last equation states, that the network manager has to find an optimal link density considering two contradictive demands: an expensive to maintain, densely connected network which integrates all the potential customers or inexpensive, sparse network which brings little income.

\begin{figure}
\centerline{\includegraphics[width=\columnwidth]{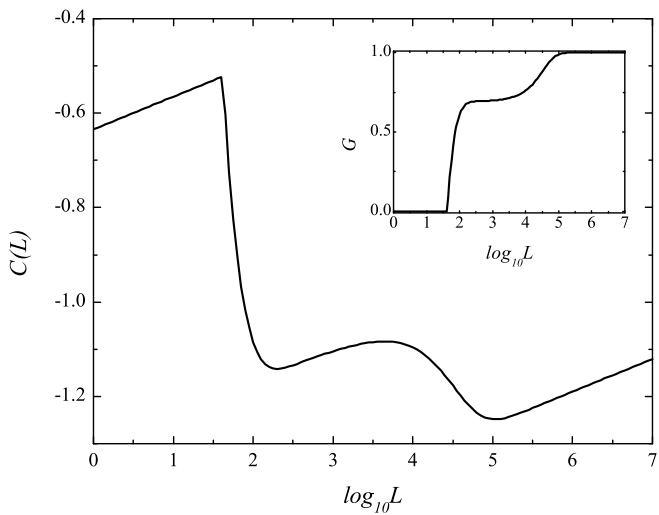}}
\caption{Cost function $C$ versus the number of global connections $L$ for the network characterized by $R=2000$, $K=20$ and $\lambda=0.97$. Inset: size of the global percolation cluster $G$, Eq.~(\ref{MFfodG}), versus $L$ for the same network.}
\label{fig6}
\end{figure}

The balance (or cost) function, Eq.~(\ref{eq:cost}), for $\lambda=0.97$ is presented in Fig.~\ref{fig6}.
During the first phase of the network construction (for $L<40$), the growing costs represent an initial investment a business owner needs to start up a firm. In the second phase one can see the two well separated minima of the cost function. They mean that the network provider who tries to operate in accordance with the economic rule (\ref{eq:cost}) has to remember that the expanding the business can lead to a temporal increase of costs and can be discouraging since one has to pass over the cost barrier.

\section{Summary}\label{sec5}

The classical blockmodel is the simplest among models of networks with community structure. Majority of scientists, who are engaged in the research of complex networks, knows this model mainly due to the fact that over the last decade, it was often used as a benchmark graph for testing community detection algorithms \cite{2002GirvanPNAS,2004NewmanPRE,2008LanciPRE}. However, in our opinion, this model deserves attention because it is also an extremely simple example of interconnected networks. For this reason, it is surprising that the percolation transition in this model has not been examined so far, although the phenomenon has been studied in a variety of much more complicated models of interconnected and multiplex networks \cite{2010ParshaniPRL,2011GaoPRL,2012EPL,2014Bianconi}.

In summary, in this paper we study percolation in the classical blockmodel. At the beginning, in Sec.~\ref{sec2}, we present a general formalism and apply it to the analytical calculation of the structural properties of classical random graphs. Next, in Sec.~\ref{sec3}, we use this formalism to obtain the self-consistent equation for the size, $G$, of the global percolation cluster in the classical blockmodel. The formalism allows to calculate the position of the phase transition (i.e. percolation threshold) at which the sample-spanning global cluster first appears. The carried out numerical simulations confirm the correctness of our theoretical predictions. Finally, in Sec.~\ref{sec4} we show, how our teoretical derivations may help to understand the cost optimization procedure in distribution networks which have a modular structure.

\section{Acknowledgments}

This work was supported by the Foundation for Polish Science (grant no.
POMOST/2012-5/5) and by the European Union within European Regional
Development Fund (Innovative Economy).

%

\end{document}